\documentclass[doublecol]{epl2}
\usepackage{color}
\usepackage{graphicx}
\usepackage{amsmath, amsthm, amssymb}
\usepackage{enumerate}
\usepackage[normalem]{ulem}

\newcommand{\be}{\begin{equation}}
\newcommand{\ee}{\end{equation}}
\newcommand{\bea}{\begin{eqnarray}}
\newcommand{\eea}{\end{eqnarray}}

\title{Stability of  columnar order  in assemblies of hard rectangles or squares}
\author{Trisha Nath\inst{1} \and Deepak Dhar\inst{2} \and R. Rajesh\inst{1}}
\institute{                    
  \inst{1} The Institute of Mathematical Sciences - C.I.T. Campus, Taramani, Chennai 600113, India\\
  \inst{2} Department of Theoretical Physics, Tata Institute of
Fundamental Research, Homi Bhabha Road, Mumbai 400005, India
}
\shortauthor{Nath \etal}

\pacs{05.50.+q}{Lattice theory and statistics}
\pacs{64.60.De}{Statistical mechanics of model systems}
\pacs{64.60.Bd}{General theory of phase transitions}

\abstract{
A system of $2\times d$ hard rectangles on square lattice  is known to 
show four different phases for $d \geq 14$. As the covered area fraction $\rho$ is increased from $0$
to $1$, the system goes from low-density disordered phase, to orientationally-ordered nematic phase,
to a columnar phase with orientational order and also broken translational invariance, to a high
density phase in which orientational order is lost.  
For large d, the threshold density for the first transition
$\rho_1^*$ tends to $0$, and the critical density for the third transition
 $\rho_3^*$ tends to $1$.  Interestingly,  simulations have  shown 
that the critical density for the second transition $\rho_2^*$ tends to a  
non-trivial finite value $\approx 0.73$, as $d \rightarrow \infty$, and $\rho_2^* \approx 0.93$ 
for $d=2$. We provide a theoretical explanation of this interesting result. We develop an approximation scheme to calculate  the surface tension between two differently ordered columnar phases. The density at which the surface tension vanishes gives an estimate    $\rho_2^* = 0.746$, for $d\to \infty$, and $\rho_2^*=0.923$ for
$d=2$. For all values of $d$, these estimates are in  good agreement with 
Monte Carlo data.
}

\begin{document}
\maketitle

\section{Introduction}

Models of particles  with  only excluded volume interaction provide very 
simple theoretical models of phase transitions. In these  models, density is the only control parameter, 
and the phase transitions  are  entropy-driven, and   geometrical in nature. These models 
serve as 
good first order models
for understanding entropy-driven transitions
in more complex systems. For instance, the liquid-solid melting transition in  
real monoatomic or colloidal solids~\cite{pusey1986} may be modeled by the hard sphere 
system that exhibits a discontinuous transition from a fluid  phase to a crystalline phase 
with increasing density~\cite{aldersphere,hoover1968}. Likewise, the qualitative features of  isotropic liquid,
nematic, smectic and solid phases observed in liquid crystals can be seen  in the much simpler models of hard rods or 
spherocylinders~\cite{degennes,onsager,bolhuis,flory,z63}. Similarly, transitions in adsorbed monolayers
(Cl on Ag) may be modeled using hard squares on lattices~\cite{taylor1985}. 
The different shapes of particles that have been 
studied include dimers~\cite{Kasteleyn1961,lieb1972,dc2008}, trimers~\cite{jesper2007}, 
hexagons~\cite{baxter}, triangles~\cite{verberkmoes}, 
squares~\cite{bellemans_nigam1nn,pears,ramola,wojciechowski,zhao,avendano}, 
discs~\cite{fernandes2007,trisha},  tetrominoes~\cite{barnes},  
rods~\cite{deepak,rrajesh,kundu2} and rectangles~\cite{kundu3,zhao2,donev}.
Despite sustained interest in hard-core models, exact solutions exist only for
hard hexagons on the triangular lattice~\cite{baxter} and 
a special limiting case of thin long cylinders in three dimensional 
continuum~\cite{onsager}.
 
In this context, the problem of hard $2 \times d$ rectangles on  a square 
lattice is specially interesting. Recently, we showed that  a system of $2 \times d$  hard
rectangles for $d \geq 14$ shows four distinct phases~\cite{kundu3}. The low-density phase is a 
disordered gas-like
phase.  On increasing density beyond a critical value $\rho_1^*(d)$,  the system enters a nematic phase,
with a larger fraction of rectangles having the same orientation (horizontal, or vertical), but with
no breaking of translational invariance. On increasing the density further, beyond a density 
$\rho_2^*(d)$, the system enters a columnar phase, where translational symmetry is broken in a 
direction perpendicular to the orientational
order: the system develops  a layered structure of period 2, with preferential occupation of particles
in either odd, or even rows, when the orientational ordering is horizontal. Finally,
beyond a density $\rho_3^*(d)$, orientational order is lost for all $d$. However, if $d$ is even,
sublattice
order is present~\cite{kundu3,kr14b,joyjit_rectangle_odd}.

There has been a lot of interest in the study of columnar, or stripe order in
recent years~\cite{edlund1}. 
It is encountered in many different settings, such as liquid crystals~\cite{degennes}, 
frustrated spin systems, both 
classical~\cite{fabien_ikhlef,asen2}, and quantum~\cite{Papanikolaou, wenzel}, and hard core
lattice gas models~\cite{bellemans_nigam1nn,ramola,trisha,kundu3,rdd2015}.
However, our understanding of  columnar ordering  is still  not very satisfactory. For instance, 
while there exists rigorous proof for the existence of the 
solid-like sublattice~\cite{peierls,dobrushin,heilmann1974phase} 
and nematic phases~\cite{dg13}, there is no corresponding result for columnar order. 

The prototypical model of columnar order is the classical $2 \times 2$  hard-square lattice 
gas, which has been studied for a long time~\cite{bellemans_nigam1nn,ramola,bellemans_nigam3nn,feng,zhi2,
slotte,lafuente2003phase,fernandes2007,rdd2015}. 
Approximation schemes like different versions of mean-field theory, cluster variational methods, 
and density functional theory,
do not capture the correlations in the columnar phase well, and the estimates of the  critical activity
$z_c(2)$ are off by a 
factor of  $5$ from the Monte Carlo results (also see Discussion). 
In this paper,  based on two different approximate calculations of interfacial tension, we obtain
$z_c(2)=52.49$ and $54.87$,   and $\rho_c(2)=0.923$ and 
$0.932$ for the critical density, to be compared with $z_c(2)\approx 97.5$, and 
$\rho_c(2) \approx 0.932$ from Monte Carlo simulations~\cite{feng,fernandes2007,zhi2}.

We study the transition between the nematic and columnar 
phases of $2 \times d$ rectangles on a square lattice, under the simplifying restriction
that  all rectangles are fully aligned (all horizontal). For  large $d$, for
most densities in the nematic/columnar range, this  approximation is very 
good (for $d=18$ and $\rho = 0.75$, the deviation from perfect  
orientational order is $\approx 0.15\%$). 
Whenever a nematic phase exists (only when $d \geq 14$), the critical threshold 
for the nematic-columnar transition  with all rectangles oriented in one direction
was shown numerically to not change much if they are allowed to 
have arbitrary orientations~\cite{trisha2}.
Also,  it is not an approximation when $d=2$, as there is no
distinction between horizontal and vertical squares.
However, restricting to only horizontal rectangles has the important 
consequence that we obtain a phase transition for all $d \geq 2$, while if we allow rectangles
of both orientations, then  systems with odd $d$ and $d \leq 9$ have  no phase 
transitions~\cite{joyjit_rectangle_odd}.

For large $d$, the estimates from Monte Carlo simulations for the 
the critical parameters of  the nematic-columnar transition are
$z_c(d) \approx 71.0 d^{-1}+O(d^{-2})$ and $\rho_c(d) \approx 0.73+0.46 d^{-1}+O(d^{-2})$~\cite{kr14b}. 
As in the case of hard squares, standard approximation techniques give poor 
theoretical estimates. 
Estimates from the singular high-density expansion in powers of ${z}^{-1/d}$ give 
$z_c(d) = 5.56 d^{-1} + O(d^{-2})$ and
$\rho_c(d) = 0.357+0.73 d^{-1}+O(d^{-2}) $~\cite{trisha2}, whereas 
a Bethe approximation gives $\rho_c(d) = 0.59+0.29 d^{-1}+O(d^{-2})$~\cite{kundu3}. 

We estimate the interfacial 
tension between two phases with different columnar order, and setting 
this to zero, we obtain a condition for the limit of stability of the columnar 
ordered phase. We have done our calculation in two different approximations. 
In the first, we  work out  the high-density expansion for the interfacial tension 
as a perturbation series  in the defect density, and estimate 
the critical point by truncating this expansion at first order. The formal power series in powers of 
defect density is a singular perturbation series 
in the activity $z$.
At the zeroth order in defect density, it is straightforward
to determine the  partition function per site in the bulk, and the interface is 
a partially directed self-avoiding walk (PDSAW). Summing over the different configurations
of the walk gives us the  zeroth order  estimates of  $z_c(d)$ and  $\rho_c(d)$. We 
then calculate the first order correction to this result due to the presence of defects in the bulk phases. 
In the second method, 
we ignore defects in the bulk, but take into account  a subset of  configurations 
of the interface with overhangs to improve the estimate of the interfacial tension.  
Both these approximations give values  [see Eqs.~\eqref{eq:zc1defect}, 
\eqref{eq:rhoc1defect}, \eqref{eq:zcoverhang} and \eqref{eq:rhocoverhang}]
that are quite close to the Monte Carlo estimates, and are a significant improvement over
earlier methods.

\section{Model and methodology}

Consider hard rectangles of size $2\times d$ ($d\geq 2$), whose
long edges are aligned along the horizontal $x$-direction, on a square lattice of size 
$N \times M$, interacting  through only excluded volume interactions (two rectangles
may not overlap). An activity $z$ is associated with each rectangle. The system is disordered
at low densities and shows columnar order at  high densities~\cite{trisha2}.
We call the phase odd (even) in which  the majority of heads (bottom
left corner) of the rectangles are in odd (even) rows.

The critical density for the nematic-columnar transition is finite
because, for 
large $d$, we can rescale the x-axis by a factor $d$, and the problem becomes equivalent to a 
model of  $1 \times 1$ oriented hard squares in  
a  space where $x$  is  continuous, but $y$ is discrete.  
Two squares cannot overlap or have a common boundary. In this continuum problem, 
the space between two lines $y=n$ and $y=n+1$ is like a continuum hard-core 
gas~\cite{tonks,takahashi}, and the transition would be expected to occur at a finite density. 
The fractional area covered  is unaffected by  rescaling of the x-axis. 
The activity  $z_{cont}$ in this continuum 
model is related to the lattice activity $z$ by $ z_{cont} = d z$.

We estimate  the critical activity $z_c(d)$ and the critical density 
$\rho_c(d)$  at the nematic-columnar transition by calculating 
within an approximation scheme,  the interfacial tension $\sigma(z)$ between the even and odd phases.
Estimating interfacial tension has been useful 
in determining the  phase diagrams of different lattice models.
Examples include Ising~\cite{muller,reed1974}, 
and Potts models~\cite{selke}.

\begin{figure}
\begin{center} 
\includegraphics[width=0.9\columnwidth]{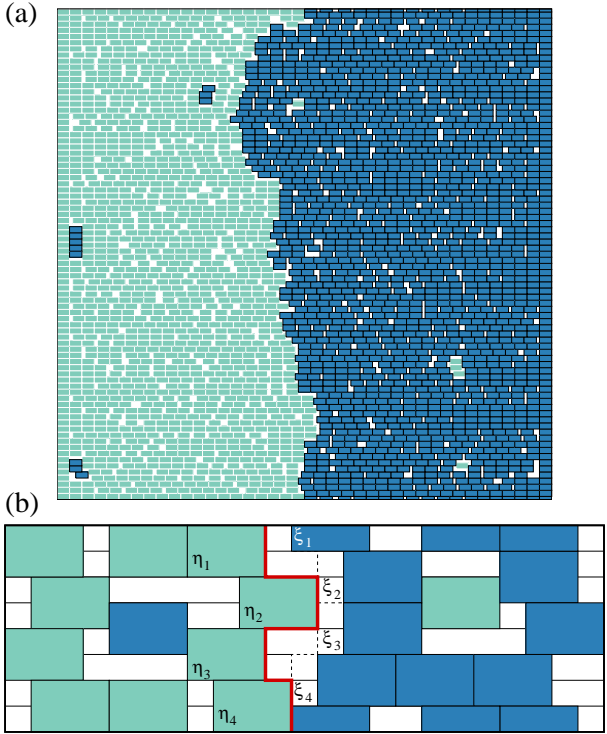}
\caption{(a) Snapshot of a typical configuration of $2 \times 4$ 
rectangles where the left (right) phase is 
constrained to be even (odd) by  fixing the rectangles at the boundary. 
Even (odd) rectangles are shown in teal (blue). The
system size is $160 \times 160$. (b) A schematic diagram of the 
interface  (thick red line) between two phases having a defect each.
$\eta_i$ is the x-coordinate of the head of the rightmost non-defect rectangle
in row $i$, and $\xi_i=\max(\eta_i,\eta_{i-1})+d$ is the minimum allowed $x$-coordinate for 
heads of odd rectangle on the $(2i -1)$th row.}
\label{fig:snap}
\end{center}
\end{figure}

To compute $\sigma(z)$, we impose an interface
by fixing the rectangles at the left (right) boundary to be even (odd). 
A typical configuration seen in a simulation of $2\times 4$ rectangles 
is
shown in Fig~\ref{fig:snap}(a). We note that there may be some empty 
space between the odd and even rectangles. To define an unique position 
of the interface for any allowed configuration of rectangles, we adopt the convention that 
the boundary between the left (even) and right
(odd) phase is placed as far left as possible. With this convention, 
there is a  well-defined interface with the bulk phases having very 
few defects (rectangles of the wrong type which when removed results in a fully ordered
columnar phase). 

Let $\Omega_{eo}(z, z_D|I)$ be the partition function with a given interface $I$,  where 
$z_D$ is the activity of a defect, and $e$ and $o$ refer to even and odd respectively.   Then, 
$\Omega_{eo}(z, z_D)$, the partition function with an interface present, is the sum of  
$\Omega_{eo}(z, z_D|I)$ over different interfaces $I$:
\be
\Omega_{eo}(z, z_D) =\sum_{I} \Omega_{eo}(z, z_D|I).
\ee
Let $\Omega_{ee}(z,z_D)$ be the partition function when all rectangles 
at the boundaries are fixed to be even (no interface). Then, 
\be
\label{sigma}
\sigma(z) =\lim_{M \rightarrow \infty} \frac{-1}{M} \ln\left[\frac{ \Omega_{eo}(z,z_D)}{\Omega_{ee}(z,z_D)}\right],
\ee
and  we define  the effective free energy $H_{\mathrm{eff}}(I)$
of any  interface $I$ to be 
\begin{equation}
H_{\mathrm{eff}}(I) = -\ln \left[ \frac{\Omega_{eo}(z,z_D|I)}{\Omega_{ee}(z,z_D)}\right].
\label{eq:heffective}
\end{equation}

In the simplest approximation, we 
ignore overhangs such that the interface is a path from top to bottom with no
upward steps allowed. Then,
the interface $I$ is uniquely defined by $\{\eta\}$, where $\eta_i$ 
is the head of the right most rectangle
of the even phase [see Fig~\ref{fig:snap}(b)]. Since the interactions are hard core, given an interface
$I$, the left  and right
phases may be occupied with rectangles independent of each other, such that
$\Omega_{eo}(z, z_D|I)=
\Omega_{L}(z,z_D|I)\Omega_{R}(z, z_D|I)$, where $L$ ($R$) refers to left (right) phase.

Each of the partition functions $\Omega_{ee}(z,z_D)$, $\Omega_{L}(z,z_D|I)$ and 
$\Omega_{R}(z, z_D|I)$, as well as the effective free energy $H_{\mathrm{eff}}(I)$
may be expanded as a perturbation series in $z_D$. For example,
$\Omega_{ee}(z,z_D)=\Omega_{ee}^{(0)}(z)+z_D\Omega_{ee}^{(1)}(z)+O(z_D^2)$,
where the superscript $(n)$ refers to the presence of $n$ defects.
Expanding Eqs.~\eqref{sigma} and \eqref{eq:heffective} in powers of $z_D$, we obtain
the zero defect contribution to $H_{\mathrm{eff}}(I)$ and $\sigma(z)$ to be
\begin{align}
&H_{\mathrm{eff}}^{(0)}(I)=-\ln\frac{\Omega_{L}^{(0)}(z|I)
\Omega_{R}^{(0)}(z|I)}{\Omega_{ee}^{(0)}(z)},
\label{heff00}\\
&\sigma^{(0)}(z)=\lim_{M\to \infty}\frac{-1}{M}\ln  \sum_I
e^{-H_{\mathrm{eff}}^{(0)}(I)}.
\label{sigma0}
\end{align}
Details of this expansion  may be found in Ref.~\cite{supplement1}.

\section{Zero defect calculation}

The calculation of $\sigma^{(0)}(z)$ is quite straightforward. In the absence of defects,
each phase consists  of only rectangles of one kind, either even or odd. This implies that
the configuration of  rectangles in a row  is independent of those in the other rows. 
Thus, the 
partition functions $\Omega_{ee}^{(0)}(z)$, $\Omega_{L}^{(0)}(z|I)$ and $\Omega_{R}^{(0)}(z|I)$ 
become products of one-dimensional partition functions, and 
we obtain  
\begin{eqnarray}
\Omega_{ee}^{(0)}(z)&=&\left[\omega_0(N)\right]^{M/2},\\
\Omega_{L}^{(0)}(z|I)&=& \prod_{i=1}^{M/2} [z \omega_0(\eta_i)], \label{eq:9}\\
\Omega_{R}^{(0)}(z|I)&=& \prod_{i=1}^{M/2}\omega_0(N-\xi_i),
\end{eqnarray}
where the interface is specified by $\{\eta\}$,
$\xi_i=\max(\eta_i,\eta_{i-1})+d$ [see Fig.~\ref{fig:snap}(b)], and  $\omega_0(N)$ is the 
partition function of a system of hard rods of length $d$ 
on a one dimensional open chain  of length $N$. The factor $z$ in Eq.~\eqref{eq:9}
accounts for the constraint
that the right most site of each even  row in the even  phase should be occupied.
It is easy to see that $\omega_0$ obeys the recursion relation 
\begin{equation}
\omega_0(\ell)= z\omega_0(\ell-d)+\omega_0(\ell-1),
\end{equation}
for $\ell=1,2,\ldots$ with $ \omega_0(0)=1$, and $ \omega_0(\ell)=0$ for $\ell <0$.
For large $N$, $\omega_0(N) = a \lambda^N [1+O(\exp(-N))]$, where $\lambda$
is the largest root of the equation
\begin{equation}
\label{eq:z}
\lambda^d-\lambda^{d-1}=z.
\end{equation}
It is easily shown that (see Ref.~\cite{supplement1})
\bea
a &=& \frac{\lambda}{d(\lambda-1)+1},
\label{eq:a_defn}\\
\sum_j \omega_0( j) x^j  &=& \frac{1}{1 - x - z x^d}.
\label{eq:gen_function}
\eea

The  effective (free) energy  of the interface $\{\eta\}$ is then
[see Eq.~\eqref{heff00}]
\be
\label{heff01}
H_{\mathrm{eff}}^{(0)} = \frac{-M}{2} \ln [ a (1 - \lambda^{-1})] +\frac{1}{2} \ln \lambda \sum_i |\eta_{i+1} - \eta_{i} |,
\ee
and we obtain  the interfacial tension when there are no defects [see Eq.~\eqref{sigma0}] to be
\be\label{sigma0value}
\sigma^{(0)}(z)=-\frac{1}{2}\ln\left[\frac{(\sqrt{\lambda}+1)^2}{d (\lambda-1)+1} \right],
\ee
where $\lambda$ is related to $z$  by Eq. (\ref{eq:z}).

The interface $\{\eta\}$ is a PDSAW, a self avoiding walk in which steps in the 
upward direction are disallowed.
From Eq.~\eqref{heff01}, we can identify the weights of steps in left, right and  down directions
to be $u=\lambda^{-1/2}$, $v=\lambda^{-1/2}$,
and $w=a (1 -\lambda^{-1})$ respectively. 
The generating function of such a PDSAW is easily seen to be
\be
G = \left[1- w - \frac{w u}{1-u} -  \frac{w v}{1-v} \right]^{-1}.
\label{eq:pdsaqgenfunct}
\ee

\begin{figure}
\begin{center} 
\includegraphics[width=\columnwidth]{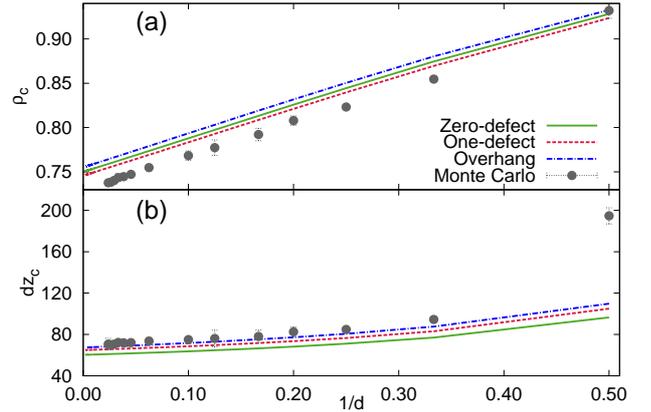}
\caption{The results obtained for (a) critical density $\rho_c$ and (b) critical activity $z_c$ 
for the disordered-columnar transition with zero defect (green line),
 one defect (red line) and overhangs (blue line) are 
compared with the results from Monte Carlo simulations (data points). The Monte Carlo
data for $d> 16$ is from Ref.~\cite{trisha2}, the data
for $d \leq 16 $ has been generated using the algorithm introduced
in Refs.~\cite{krds12,kundu2}, and the data for $d=2$ is from Refs.~\cite{feng,fernandes2007,zhi2}.
}
\label{fig:zcrhoc}
\end{center} 
\end{figure}

The phase transition occurs when the interfacial tension is zero or equivalently when the generating
function $G$ in Eq.~\eqref{eq:pdsaqgenfunct} diverges. Setting $\sigma^{(0)}(z)=0$, we obtain 
the estimates of the critical parameters $\lambda_c^*(d)$ and $\rho_c^*(d)$  to be  
\bea
\sqrt{\lambda_c^*(d)}&=&\frac{1+\sqrt{1-d+d^2}}{d-1},\\
\rho_c^*(d)&=&\frac{d(1+d+2\sqrt{1-d+d^2})}{(d+\sqrt{1-d+d^2})^2},
\eea
and $z_c^*(d)=\lambda_c^*(d)^{d}-\lambda_c^*(d)^{d-1}$.
For hard-squares ($d=2$), $\lambda_c^*(2)=(\sqrt{3}+1)^2$, 
$z_c^*(2)=(4+2\sqrt{3})(3+2\sqrt{3})\approx48.25$, 
$\rho_c^*(2)=(6+4\sqrt{3})(7-4\sqrt{3})\approx0.928$. These  
should be compared with numerically obtained $z_c=97.5$ and $\rho_c=0.932$~\cite{feng,fernandes2007,zhi2}.
For large $d$ limit, the solution is  $\lambda_c^*=1+3 d^{-1}+O(d^{-2})$, $z_c=60.26 d^{-1}+O(d^{-2})$, 
and $\rho_c^*=0.75+0.375 d^{-1}+O(d^{-2})$. These should be 
compared with the numerically obtained values $z_c \approx71.0/d$ and 
$\rho_c \approx 0.73+0.46/d$~\cite{kr14b,trisha2}. For intermediate values of $d$,
the analytical expression is compared with the Monte Carlo results 
in Fig.~\ref{fig:zcrhoc}. We note that  the results of calculation
with no defects  are  already  significantly better than earlier estimates.

\section{One defect calculation}

\begin{figure}
\begin{center} 
\includegraphics[width=0.9\columnwidth]{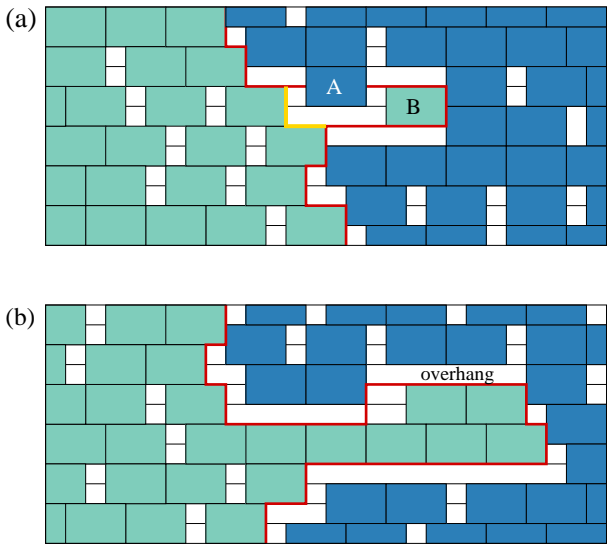}
\caption{(a) A configuration with a single defect where the interface  may be drawn (red or yellow) 
so that the defect is an odd rectangle (A) in the even phase,   or an even rectangle (B) in the odd phase.
(b) A configuration in which  the
interface has  one  overhang. 
} 
\label{fig:1def}
\end{center} 
\end{figure}

Consider now the corrections to the interfacial tension to the  first order in $z_D$.  
For this, we need to compute $\Omega_{ee}^{(1)}$, 
$\Omega_{L}^{(1)}(z|I)$, and $\Omega_{R}^{(1)}(z|I)$, the partition functions when
a single defect is present.  There are configurations which may be considered as single odd defect in the even phase, or equally as a single even defect in the odd phase by  repositioning the interface [see Fig.~\ref{fig:1def}(a)].   To prevent  double counting, we adopt the convention  that a defect in the  even  phase 
should have at least two  even
rectangles to its right. Else, we  redefine the interface locally.

A defect rectangle  affects the occupation of utmost the two rows adjacent to it,
by splitting the row(s) 
into two intervals that may be occupied independent of each other. However, unlike the zero defect
case, the intervals may be small instead of being extensive in $N$, and therefore it is necessary to
know $\omega_0(\ell)$ for all $\ell$ and not just for large $\ell$. The detailed calculation of the
contribution from one defect may be found in Ref.~\cite{supplement1}. 
The one defect
calculation reduces to finding the expectation values of two properties of the PDSAW: the mean step 
length and two-point correlations between
adjacent steps.

By equating $\sigma(z)=\sigma^{(0)}(z)+z_D\sigma^{(1)}(z)$ to zero we 
obtain 
\bea
z_{c}^*(d) &=& 65.23 d^{-1}+ O(d^{-2}), \label{eq:zc1defect}\\
\rho_{c}^*(d)&=& 0.746+0.374 d^{-1} + O(d^{-2}).  \label{eq:rhoc1defect}
\eea
For hard squares, we obtain $z_c^*(2) = 52.49$ and $\rho_c^*(2) = 0.923$. These results and those
for general $d$ are compared with the Monte Carlo results as well as the contribution from zero
defects in Fig.~\ref{fig:zcrhoc}. The one defect calculation improves the estimates for $z_c(d)$ and
$\rho_c(d)$ for all $d>2$. For $d=2$, while $z_c(2)$ increases towards the Monte Carlo results, $\rho_c(2)$
decreases slightly away from the Monte Carlo results.

\section{Calculation including overhangs}

Another improvement over the zero defect calculation would be to include in the sum over interfaces, interfaces with overhangs. 
In this calculation, we assume perfect columnar order in each of the phases to the left or right of interface.
We consider all interfaces in which the overhangs have  height at most $1$. 
These interfaces may be defined as self-avoiding walks such that if the walk once reaches the  
layer $y$ it never goes to the layer  $y-4$ any time later, but may go back to 
$y-2$, a single vertical step of the walk being of length two lattice spacings. An interface with one overhang  of height 1  is shown in  Fig.~\ref{fig:1def}(b).

If we denote the up, right, down, and left steps of the walk by letters $U$, $R$, $D$, and $L$ respectively, 
it is easily seen that an overhang configuration consists of sequence of steps of the type $R^aUR^bD$, where $a,b >d$.  Define
\bea
W_R &=& \sum_{a,b \geq d}^{\infty} R^a U R^b D,\\
W_R^* &=& W_R + W_R W_R + W_R W_R W_R+ \ldots,
\eea
and a similar definition for sum over overhang configurations to the left $W_L^*$.  $W_R^*$ and
$W_L^*$ generate overhangs of height $1$.

It is straight forward to write down the sum over configurations of rectangles corresponding to an interface with such  overhangs as a product of one-dimensional partition functions 
$\omega_0$.  Note that in the definition of $W_R$, there is  a  sum over  lengths of steps $a$ and $b$.  
These come with a weight $\lambda^{-3(a+b)/2}$. This implies that the corresponding weight is expressed 
in terms of  the function $\sum_i \omega_0(i) \lambda^{-3 i/2}$, which is a simple rational function of 
$\lambda^{-3/2}$ and $z$ [see Eq.~\eqref{eq:gen_function}], and this simplifies the expressions of weights of $W_R$ and $W_R^*$ enormously. Details of the calculation are given in Ref.~\cite{supplement1}. Equating the surface tension to zero, we obtain the estimate 
\bea
z_c^{**}(d) &=& 67.78 d^{-1}+ O(d^{-2}), \label{eq:zcoverhang}\\
\rho_c^{**} (d) &=& 0.755+0.382 d^{-1} + O(d^{-2}), \label{eq:rhocoverhang}
\eea
and
$z_c^{**}(2) = 54.87$ and $\rho_c^{**}(2) = 0.9326$.  These are an improvement over the estimates
of $z_c(d)$ from the one defect calculations [see Fig.~\ref{fig:zcrhoc} for comparison]. However,
the calculation of overhangs does not affect the bulk quantities, and 
therefore the estimates for critical density  $\rho_c(d)$
increase. While the estimates  for  $\rho_c(d)$ matches with the Monte Carlo result for 
$d=2$, it increases away
from the Monte Carlo results for $d >2$.

\section{Discussion}

In this paper,  we  determined the  critical 
activity and critical density   of transition between the nematic and columnar phases 
in the problem of $2 \times d$ oriented  rectangles  on  a square lattice approximately, 
by estimating the interfacial tension 
between two differently ordered columnar phases. This was done in two ways: first 
in terms of a perturbation series in number of defects, and  second by including 
a subset of overhangs in the interface. Keeping only the first order terms in the expansion in the first case
and overhangs of height $1$ in the second case,   we obtain estimates that  are in fair 
agreement with Monte 
Carlo results and an improvement over earlier estimates.
For example, when $d=2$,  Monte Carlo estimates are $z_c(2) \approx 97.5$, and $\rho_c(2) \approx 0.932$, while we obtain $z_c(2)=52.49$ and $54.87$,   and $\rho_c(2)=0.923$ and 
$0.932$. These should be compared with estimates from high density expansion 
($z_c(2) =14.86$)~\cite{bellemans_nigam1nn,ramola},
a lattice version of density functional theory 
($z_c(2) = 11.09$)~\cite{lafuente2003phase,lafuente2002phase}
an approximate counting method using transfer matrices ($z_c(2) = 11.09$)~\cite{temperley},
a Flory type mean-field theory (with similar prediction)~\cite{fernandes_sq}, and
cluster variational method ($z_c(2)=17.22$)~\cite{bellemans_nigam1nn}.
A calculation of interfacial tension 
in the next nearest neighbour 
Ising antiferromagnet,  gives a better estimate $z_c(2)=135.9$~\cite{slotte}, still off  by about $40\%$. 
For the critical density, the high density expansion gives  
$\rho_c(2)=0.75$, cluster variation method gives $\rho_c(2)=0.80$, and the 
transfer matrix calculations $\rho_c(2)=0.76$. 
Likewise,  we obtain significantly better estimates for the critical parameters for 
the  hard rectangle gas also.
These  may be systematically
improved by including more defects or overhangs.
Also, the calculations may be extended to a system of 
rectangles of same width and different lengths\footnote{In this case, the zero defect
calculation gives $\rho_c^{(i)}/d_i+ 2 \sqrt{1-\sum_i \rho_c^{(i)}}=1$, where $d_i$ are the different
lengths.}.  Similar results may be derived for mixtures of squares and dimers~\cite{rdd2015}.

The interfacial tension was calculated for a system with only horizontal rectangles.
However, it is straightforward to
show that for large $d$, adding vertical rectangles introduces only exponentially small (in $d$)
corrections to the
critical density. For $d=2$, this approximation is irrelevant. However, for small $d>2$, the effects are 
not negligible~\cite{kundu3,joyjit_rectangle_odd}.

On general grounds, the nematic-columnar transition is expected to be in the Ising 
universality class~\cite{kundu3,joyjit_rectangle_odd}. 
In our calculation, the interfacial tension varies as $|z-z_c|$ near the transition point, consistent 
with the behavior in the Ising universality class~\cite{mccoy2014}.

\begin{acknowledgments}
The simulations were done on the supercomputer
Annapurna at the Institute of Mathematical Sciences. 
DD was supported in part by the Department of Science and Technology (India) 
under the grant   DST-SR/S2/JCB-24/2005.
\end{acknowledgments}

\bibliographystyle{eplbib}

 \end{document}


\title{{\it Supplemental Information for} Stability of columnar order in assemblies of hard rectangles
or squares}
\author{Trisha Nath}
\author{Deepak Dhar}
\author{R. Rajesh}

\maketitle

\section{\label{A}Expressions for one-defect contribution to $H_{\mathrm{eff}}(I)$ and $\sigma(z)$}

We Taylor expand $\Omega_{ee}(z,z_d)$ in powers of $z_d$, and write
\be
\Omega_{ee}(z,z_d) = \Omega_{ee}^{(0)}(z) + z_d \Omega_{ee}^{(1)}(z) +z_D^2 \Omega_{ee}^{(2)}(z) + \ldots,
\ee
and similar expressions for $\Omega_L(z|I)$, $\Omega_R(z|I)$ etc. Using the definition of $H_{\mathrm{eff}}(I)$ from Eq. (3) of main text, and expanding each partition functions in powers of $z_d$, we obtain 
\begin{align}
&H_{\mathrm{eff}}^{(1)}(I) = 
\frac{\Omega_{ee}^{(1)}(z)}{\Omega_{ee}^{(0)}(z)}-
\frac{\Omega_{L}^{(1)}(z|I)} {\Omega_{L}^{(0)}(z|I)}-
\frac{\Omega_{R}^{(1)}(z|I)}{\Omega_{R}^{(0)}(z|I)},\label{heff11}\\
&\sigma^{(1)}(z)=\lim_{M\to \infty}\frac{1}{M}
\frac{\sum\limits_I
H_{\mathrm{eff}}^{(1)}(I) e^{-H_{\mathrm{eff}}^{(0)}(I)}}{\sum\limits_I
e^{-H_{\mathrm{eff}}^{(0)}(I)}}.
\label{sigma1}
\end{align}

\section{\label{B} Derivation of Eq. (11)}

We start with Eq. (10) of main text.  This equation  has $d$ roots. Let them be denoted by $\lambda_i$, $i=1,\,2,\ldots,d$ where $\lambda_1$, 
 the largest root, will be denoted by $\lambda$. Then for any $\ell$, the one-dimensional partition function can thus be written as
 \be\label{omegaee0full}
 \omega_0(\ell)=\sum\limits_{i=1}^{d}a_i\lambda_i^{\ell}.
 \ee
For large $\ell$, we can approximate $\omega_0(\ell)\approx a\lambda^{\ell}$, where $a$ is $a_1$ in Eq.~\eqref{omegaee0full}.
 
 We calculate $a$ from the density of occupied sites. Suppose $\rho(d)$ is the density of the rectangles on the one-dimensional chain of length $\ell$. Then $\rho(d)$ is given by
\be\label{rhodef1}
\rho(d)=\frac{d}{\ell}\,z\frac{d}{dz}\ln \omega_o(\ell).
\ee
 On substitution of $\omega_0(\ell)\approx a\lambda^{\ell}$ we obtain 
 \be\label{rho1}
 \rho(d)=\frac{d(\lambda-1)}{d(\lambda-1)+1}.
 \ee
 Consider a site $m$ in the bulk of the chain. The probability that it is empty is 
 $ [\omega_0(m-1) \omega_0(1) \omega_0(\ell-m)]/\omega_0(\ell)$. Clearly this is equal to $1-\rho(d)$.  Substituting $\omega_0(\ell)\approx  a\lambda^{\ell}$ we obtain,
\be\label{rho2}
1-\rho(d)=\frac{a}{\lambda}.
\ee
Substituting $\rho(d)$ from Eq.~\eqref{rho1}, we obtain $a$ as
\be\label{aa}
a=\frac{\lambda}{d(\lambda-1)+1}.
\ee
 
\section{\label{C}Calculation of $\sigma^{(1)}(z)$ }
  
\begin{figure}
\includegraphics[width=0.7\columnwidth]{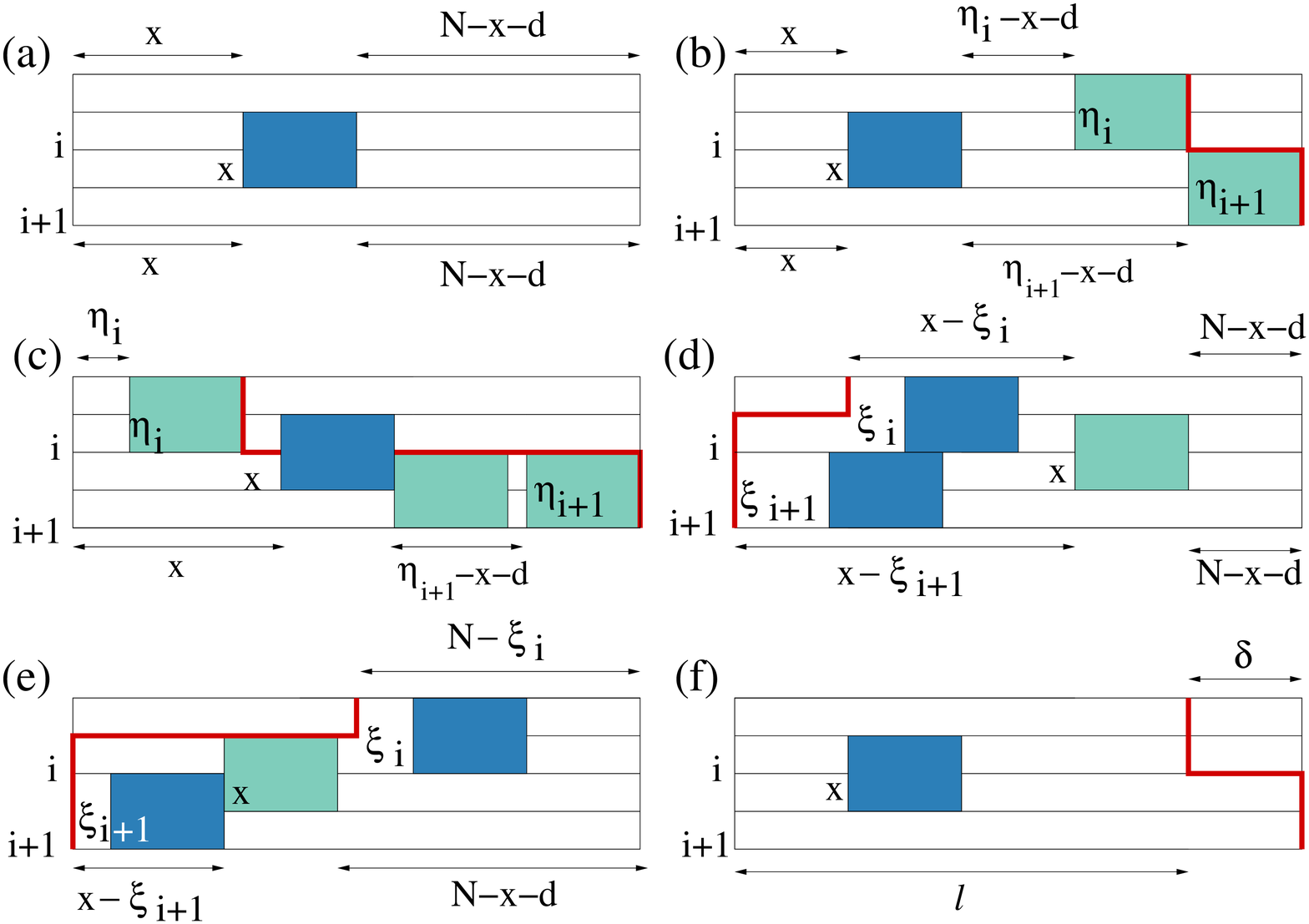}
\caption{Examples of single defects for 
(a) $\Omega_{ee}^{(1)}$, 
(b)-(c) $\Omega_{eo,L}^{(1)}$
and (d)-(e) $\Omega_{eo,R}^{(1)}$. (f) A generalized two-row system with a single defect in the bulk. 
The red lines denote the position of the interface.} 
\label{fig:S1def}
\end{figure}

Consider the contribution to the partition function when a single defect is present.  Let the defect 
rectangle be placed such that it affects the occupation of rows $i$ and $i+1$, where $i=1,\ldots, M/2$,
with the $x$-coordinate of the head being denoted by $x$. Examples of defects for different boundary 
conditions
are shown in Fig.~\ref{fig:S1def}(a)-(e). To avoid over-counting, we impose the constraints that a 
defect in the left phase should at least have two even rectangles to its right [see Fig.~\ref{fig:S1def}(b)-(c)]
and a defect in the right phase should have at least one odd rectangle to its 
left [see Fig.~\ref{fig:S1def}(d)-(e)]. 
In each of the 
cases shown in Fig.~\ref{fig:S1def}(a), (b), and (d),  the defect splits the rows $i$ and $i+1$ into 
two intervals, 
the occupation of each interval being 
independent of the other. However in configurations of type shown in Fig.~\ref{fig:S1def}(c), and (e), 
the defect affects only one of the rows, splitting it into two open intervals.
Thus, we
obtain 
\begin{align}
&\frac{\Omega_{ee}^{(1)}}{\Omega_{ee}^{(0)}}= \frac{M}{2}\sum\limits_{x=0}^{N-d}\left[\frac{\omega_0(x)
\omega_0(N-x-d)}{\omega_0(N)}\right]^2,\label{omegaee3}\\
&\frac{ \Omega_{eo,L}^{(1)}}{\Omega_{eo,L}^{(0)}}= \sum_{i=1}^{M/2}\left[\sum_{x=0}^{x_L}
\frac{\omega_0^2(x)\omega_0(\eta_i-x-d) \omega_0(\eta_{i+1}-x-d)}
{\omega_0(\eta_i)\omega_0(\eta_{i+1})}+\sum\limits_{x=x_L+2d}^{x_L'-2d}
\frac{\omega_0(x)[\omega_0(x_L'-x-d)-1]}{\omega_0(x_L')}\Theta(|\Delta\eta_i|-3d)
\right], \label{omega11l} \\
&\frac{\Omega_{eo,R}^{(1)}}{\Omega_{eo,R}^{(0)}}=\sum_{i=1}^{M/2}  \Bigg[ \sum_{x=x_R-d}^{N-d}
\frac{\omega_0^2(N-x-d) [\omega_0(x-\xi_i)\omega_0(x-\xi_{i+1})-1]}{ \omega_0(N-\xi_i)\omega_0(N-\xi_{i+1})}
\nonumber\\
&+ \sum_{x=x_R''}^{x_R-d-1}
\frac{ \omega_0^2(N-x-d)[\omega_0(x-x_R')-1]  \Theta(|\Delta\xi_i|-d-1)}{\omega_0(N-\xi_i)\omega_0(N-\xi_{i+1})}+ \sum_{x=x_R'+d}^{x_R-2d}
\frac{ \omega_0(N-x-d)[\omega_0(x-x_R')-1] \Theta(|\Delta\xi_i|-2d)}
{\omega_0(N-x_R')}
\Bigg],
\label{omega11r}
\end{align}
where $x_L=\min(\eta_i,\eta_{i+1})-d$, $x_L'=\max(\eta_i,\eta_{i+1})$, $x_R=\max(\xi_i,\xi_{i+1})+d$, $x_R'=\min(\xi_i,\xi_{i+1})$, 
$x_R''=\max(x_R-2d+1,x_R'+d)$, and $\Theta(x)$ is a step function
\begin{equation} 
\Theta(x)= 
\begin{cases}
    1,& \text{if } x\geq 0,\\
    0,              & \text{otherwise},
\end{cases}
\end{equation}
and we have dropped
the dependence on $z,I$ for notational simplicity. The first and second terms 
of Eq.~\eqref{omega11l} are 
contributions from Fig.~\ref{fig:S1def}(b) and (c) respectively. 
Similarly, the first term of Eq.~\eqref{omega11r} is contribution from Fig.~\ref{fig:S1def}(d), whereas the 
rest are contributions from configurations of the type Fig.~\ref{fig:S1def}(e).

In the sums, when $x$ is not close to the  lower or upper limits, we may substitute 
$\omega_0(\ell) \approx a \lambda^\ell$, else there are corrections due to the boundaries. To
characterize these edge effects, we consider the partition function of a two-row system
with a single defect and boundary as shown in Fig.~\ref{fig:S1def}(f). The partition
function of this shape for large $\ell$ is
\begin{align}
\sum_{x=0}^{\ell -d}
\frac{\omega_0^2(x)\omega_0(\ell-x-d) \omega_0(\ell+\delta-x-d)}
{\omega_0(\ell)\omega_0(\ell+\delta)}=\frac{a^2}{\lambda^{2d}}[(\ell-d+1)+f(0)+f(\delta)],
\label{feta2}
\end{align}
where $f(\delta)$ encodes the edge-effects,
\begin{align}
 f(\delta)=\left[2\sum\limits_{p=2}^d \frac{a_p}{a}\frac{1}{1-\frac{\lambda_p}{\lambda}}\left(1+\left(\frac{\lambda_p}{\lambda}\right)^{|\delta|}\right)+\sum\limits_{p,q=2}^d \frac{a_pa_q}{a^2}\frac{1}{1-\frac{\lambda_p\lambda_q}{\lambda^2}}\left( \frac{\lambda_p}{\lambda}\right)^{|\delta|}\right].\label{feta3}
\end{align}

Equations.~\eqref{omegaee3},~\eqref{omega11l} and \eqref{omega11r}
may thus be simplified to
\begin{align}
&\frac{\Omega_{ee}^{(1)}}{\Omega_{ee}^{(0)}}= \frac{M}{2} \frac{a^2}{\lambda^{2d}}[(N-d+1)+2f(0)],
\label{omegaee4}\\
&\frac{ \Omega_{eo,L}^{(1)}}{\Omega_{eo,L}^{(0)}}=\frac{a^2}{\lambda^{2d}}
\sum\limits_{i=1}^{M/2} \left[x_L+1+f(0)+f(\Delta\eta_i)+p(|\Delta\eta_i|-3d)\right],
\label{omega12l}\\
&\frac{ \Omega_{eo,R}^{(1)}}{\Omega_{eo,R}^{(0)}}=\frac{a^2}{\lambda^{2d}}
\sum_{i=1}^{M/2}\left[N\!-\!x_R\!+\!1\!+\!f(0)+f(\Delta\xi_{i})
-\frac{a^{-2} \lambda^{-|\Delta\xi_i|}}{1-\lambda^{-2}}+p(|\Delta\xi_i|-2d)+q(|\Delta\xi_i|)\right],
\label{omega12r}
\end{align}
where $\Delta\xi_{i}=\xi_{i+1}-\xi_{i}$, and $p(x)$ and $q(x)$ are defined as
\begin{align}
 p(x)&=\left[\frac{\lambda^d}{a}(x+1)-\frac{1-\lambda^{-x-1}}{a^2(1-\lambda^{-1})} 
 +\sum\limits_{p=2}^d\frac{a_p\lambda_p^d }{a^2(1-\frac{\lambda_p}{\lambda})}  \left(1-  \left( \frac{\lambda_p}{\lambda}\right)^{x+1}\right) \right]\Theta(x),\label{px}\\
  q(x)&=\frac{\Theta(x-d-1)}{a(1-\lambda^{-1})}\left[-1+\frac{\lambda^{-x}}{a(1+\lambda^{-1})}+ \sqrt{\lambda}^{x-|x-2d+1|-1} 
  -\frac{\lambda^{-|x-2d+1|-1}}{a(1+\lambda^{-1})}\right.\nonumber\\
  &\left.+\sum\limits_{p=2}^d\frac{a_p}{a}\frac{1-\lambda^{-1}}{1-\frac{\lambda_p}{\lambda^2}}\left( \frac{\lambda_p}{\lambda}\right)^{x}\left( \left( \frac{\lambda}{\lambda_p}\right)^{x-
  |x-2d+1|-1}-1\right)\right].\label{qx}
\end{align}

The contribution from a single defect to the free energy of the interface, $H_{\mathrm{eff}}^{(1)}(I)$ 
[see Eq.~\eqref{heff11}]
may now be evaluated from Eqs.~\eqref{omegaee4}, \eqref{omega12l} and \eqref{omega12r}:
\begin{align}\label{heff12}
H_{\mathrm{eff}}^{(1)}(I)&=\frac{a^2}{\lambda^{2d}}\sum_{i=1}^{M/2}
\Big[2d -1+\frac{2 |\Delta\eta_i|+|\Delta\xi_{i}|}{2}+\frac{a^{-2} \lambda^{-|\Delta\xi_i|}}{1-\lambda^{-2}}-f(\Delta\eta_i)-p(|\Delta\eta_i|-3d)-f(\Delta\xi_{i})\nonumber\\
&-p(|\Delta\xi_i|-2d)-q(|\Delta\xi_i|)\Big].
\end{align}
Substituting Eq.~\eqref{heff12} in Eq.~\eqref{sigma1} we obtain interfacial tension with one defect as
\begin{align}
 \sigma^{(1)}(z)&=\frac{a^2}{M\lambda^{2d}}\sum\limits_{i=1}^{M/2} \left[2d-1+
 \frac{\sqrt{\lambda}-1}{\sqrt{\lambda}+1}
 \sum\limits_{\Delta\eta_i}(|\Delta\eta_i|-f(\Delta\eta_i)-p(|\Delta\eta_i|-3d))\sqrt{\lambda}^{-|\Delta\eta_i|} \right.\nonumber\\
 &\left.+ \left(\frac{\sqrt{\lambda}-1}{\sqrt{\lambda}+1}\right)^2
 \sum\limits_{\Delta\eta_i,\Delta\eta_{i+1}}\left(\frac{|\Delta\xi_i|}{2}-f(\Delta\xi_{i})
+\frac{a^{-2} \lambda^{-|\Delta\xi_i|}}{1-\lambda^{-2}}-p(|\Delta\xi_i|-2d)-q(|\Delta\xi_i|) \right)\sqrt{\lambda}^{-|\Delta\eta_i|-|\Delta\eta_{i+1}|}
 \right].\label{sigmafinal1}
\end{align}

Summing over $\Delta\eta$ and $\Delta\xi$s, each of the terms in Eq.~\eqref{sigmafinal1} are evaluated 
to be
\begin{align}
  \frac{\sqrt{\lambda}-1}{\sqrt{\lambda}+1} \sum\limits_{\Delta\eta_i}
  |\Delta\eta_i|\sqrt{\lambda}^{-|\Delta\eta_i|}&= \frac{2\sqrt{\lambda}}{\lambda-1},\label{eq1}\\
  \left(\frac{\sqrt{\lambda}-1}{\sqrt{\lambda}+1}\right)^2 \sum\limits_{\Delta\eta_i,\Delta\eta_{i+1}}
  \frac{|\tilde{\xi_i}|}{2}\sqrt{\lambda}^{-|\Delta\eta_i|-|\Delta\eta_{i+1}|}&=\frac{\sqrt{\lambda}+\lambda+\lambda^{3/2}}{(\sqrt{\lambda}-1)(\sqrt{\lambda}+1)^3},\label{eq2}\\
 \frac{a^{-2}}{1-\lambda^{-2}} \left(\frac{\sqrt{\lambda}-1}{\sqrt{\lambda}+1}\right)^2 \sum\limits_{\Delta\eta_i,\Delta\eta_{i+1}}
  \sqrt{\lambda}^{-2|\Delta\xi_{i}|-|\Delta\eta_i|-|\Delta\eta_{i+1}|}&=\frac{\lambda^2}{a^2(\sqrt{\lambda}-1)(\sqrt{\lambda}+1)^3},\label{eq3}\\
 \frac{\sqrt{\lambda}-1}{\sqrt{\lambda}+1} \sum\limits_{\Delta\eta_i}  f(\Delta\eta_i) \sqrt{\lambda}^{-|\Delta\eta_i|}&=2\sum\limits_{p=2}^d\frac{a_p(\lambda^2-\lambda_p)}{a(1-\frac{\lambda_p}{\lambda})(\sqrt{\lambda}+1)(\lambda^{3/2}-\lambda_p)}\nonumber\\
 &+\sum\limits_{p,q=2}^d \frac{a_pa_q(\sqrt{\lambda}-1)(\lambda^3-\lambda_p\lambda_q)}{a^2(1-\frac{\lambda_p\lambda_q}{\lambda^2})(\sqrt{\lambda}+1)(\lambda^{3/2}-\lambda_p)(\lambda^{3/2}-\lambda_q)},\label{eq4}
   \end{align}
 \begin{align}
 & \left(\frac{\sqrt{\lambda}-1}{\sqrt{\lambda}+1}\right)^2 \sum\limits_{\Delta\eta_i,\Delta\eta_{i+1}}  
 f(\Delta\xi_{i})\sqrt{\lambda}^{-|\Delta\eta_i|-|\Delta\eta_{i+1}|}=2\sum\limits_{p=2}^d\frac{a_p(2\lambda^2+2\lambda^{5/2}+\lambda^3-\lambda_p(1+2\sqrt{\lambda}+2\lambda))}{a(1-\frac{\lambda_p}{\lambda})(\sqrt{\lambda}+1)^3(\lambda^{3/2}-\lambda_p)}\nonumber\\
&\qquad\qquad\qquad\qquad +\sum\limits_{p,q=2}^d \frac{a_pa_q(\lambda^3(\lambda^{3/2}+\lambda+\sqrt{\lambda}-1)-\lambda^2(\lambda_p+\lambda_q)(1+\sqrt{\lambda})+\lambda_p\lambda_q(1+\sqrt{\lambda}+\lambda-\lambda^{3/2}))        )}{a^2(1-\frac{\lambda_p\lambda_q}{\lambda^2})(\sqrt{\lambda}+1)^3(\lambda^{3/2}-\lambda_p)(\lambda^{3/2}-\lambda_q)},\label{eq5}
\end{align}
  \begin{align}
 &\frac{\sqrt{\lambda}-1}{\sqrt{\lambda}+1} \sum\limits_{\Delta\eta_i}
 p(|\Delta\eta_i|-3d))\sqrt{\lambda}^{-|\Delta\eta_i|}=\frac{2\lambda^{(1-3d)/2}(a(1-3d)\lambda^d-\frac{\lambda}{\lambda-1})}{a^2(\sqrt{\lambda}+1)}+\frac{2\lambda^{(1-d)/2}(1+3d(\sqrt{\lambda}-1))}{a(\lambda-1)}\nonumber\\
 &\qquad\qquad\qquad\qquad\qquad\qquad+\frac{2\lambda^{-3(d-1)/2}}{a^2(\sqrt{\lambda}+1)^2(\lambda^{3/2}-1)}
 +2\sum\limits_{p=2}^d\frac{a_p\lambda^{-3(d-1)/2}\lambda_p^d}{a^2(\lambda-\lambda_p)(\sqrt{\lambda}+1)}\left(1 
 -\frac{(\sqrt{\lambda}-1)\lambda_p}{\lambda^{3/2}-\lambda_p}\right),\label{eq6}
 \end{align}
  \begin{align}
& \left(\frac{\sqrt{\lambda}-1}{\sqrt{\lambda}+1}\right)^2 \sum\limits_{\Delta\eta_i,\Delta\eta_{i+1}} 
 p(|\Delta\xi_i|-2d)\sqrt{\lambda}^{-|\Delta\eta_i|-|\Delta\eta_{i+1}|}=\frac{2\lambda^{(1-2d)/2}(1+\sqrt{\lambda}+\lambda)(a(1-2d)\lambda^d-\frac{\lambda}{\lambda-1})}{a^2(\sqrt{\lambda}+1)^3}\nonumber\\
 &\qquad\qquad\qquad\qquad\qquad\qquad\qquad\qquad+\frac{2\lambda^{(3-2d)/2}}{a^2(\sqrt{\lambda}-1)(\sqrt{\lambda}+1)^4}+\frac{2\sqrt{\lambda}(1+\sqrt{\lambda}+\lambda)(1+2d(\sqrt{\lambda}-1) ) }{a(\sqrt{\lambda}+1)^3(\sqrt{\lambda}-1)}\nonumber\\
 &\qquad\qquad\qquad\qquad\qquad\qquad\qquad\qquad+2\sum\limits_{p=2}^d\frac{a_p\lambda^{(3-2d)/2}\lambda_p^d(1+\sqrt{\lambda}+\lambda)}{a^2(\lambda-\lambda_p)(\sqrt{\lambda}+1)^3}\left(1 
 -\frac{(\sqrt{\lambda}-1)\lambda_p}{\lambda^{3/2}-\lambda_p}\right),\label{eq7}
 \end{align}
  \begin{align}
 &\left(\frac{\sqrt{\lambda}-1}{\sqrt{\lambda}+1}\right)^2 \sum\limits_{\Delta\eta_i,\Delta\eta_{i+1}} 
 q(|\Delta\xi_i|)\sqrt{\lambda}^{-|\Delta\eta_i|-|\Delta\eta_{i+1}|}= 
 \frac{2 \lambda ^{2-\frac{3 d}{2}}}{a^2 \left(\sqrt{\lambda }+1\right)^3 \left(\lambda ^2-1\right)}
 -\frac{2 \left(\lambda +\sqrt{\lambda }+1\right) \lambda ^{1-\frac{d}{2}}}{a \left(\sqrt{\lambda }-1\right) \left(\sqrt{\lambda }+1\right)^4}\nonumber\\
 &\qquad\qquad\qquad-\frac{2 \lambda ^{1-\frac{3 d}{2}} \left(2 \lambda ^{\frac{d}{2}+1}+3 \lambda ^{\frac{d}{2}+2}+\lambda ^{\frac{d}{2}+3}+\lambda ^{\frac{d+1}{2}}+3 \lambda ^{\frac{d+3}{2}}+2 \lambda ^{\frac{d+5}{2}}-\lambda ^{3/2}-3 \lambda ^{5/2}-\lambda ^{7/2}-2 \lambda ^3-2 \lambda ^2\right)}{a^2 \left(\sqrt{\lambda }+1\right)^4 \left(\lambda ^{3/2}+\lambda ^{5/2}-\lambda -1\right)}\nonumber\\
 &\qquad\qquad\qquad\qquad+\frac{2 \left(-2 \lambda ^{2-\frac{d}{2}}-2 \lambda ^{3-\frac{d}{2}}-\lambda ^{\frac{3}{2}-\frac{d}{2}}-3 \lambda ^{\frac{5}{2}-\frac{d}{2}}-\lambda ^{\frac{7}{2}-\frac{d}{2}}+5 \lambda ^{3/2}+3 \lambda ^{5/2}+\lambda ^3+5 \lambda ^2+3 \lambda +\sqrt{\lambda }\right)}{a \left(\sqrt{\lambda }+1\right)^4 \left(\lambda ^{3/2}-1\right)}\nonumber\\
 &-2\sum\limits_{p=2}^d\frac{a_p\lambda^2}{a^2(\lambda^2-\lambda_p)}\frac{\lambda^{-3d/2}(\lambda^{\frac{3}{2}}-1)\lambda_p^{d+1}}{(\sqrt{\lambda}+1)^3(\lambda^{\frac{3}{2}}-\lambda_p)}\nonumber\\ 
 &+2\sum\limits_{p=2}^d\frac{a_p\lambda^2}{a^2(\lambda^2-\lambda_p)}\Bigg[\lambda^{-1-2d}\Bigg(
 \lambda ^3 \lambda_p(\sqrt{\lambda } \sqrt{\lambda_p})^{2 d-1}-\lambda\lambda_p^{3/2} \left(\sqrt{\lambda } \sqrt{\lambda_p}\right)^{2 d} -2 \lambda ^{\frac{d}{2}+3} \lambda_p^{d+\frac{1}{2}}-2 \lambda ^{\frac{d}{2}+4} \lambda_p^{d+\frac{1}{2}}-\lambda^{d+2} \lambda_p^{d+\frac{1}{2}}\nonumber\\
 &+2 \lambda^{d+\frac{5}{2}} \lambda_p^{d+\frac{1}{2}}+2 \lambda^{d+3} \lambda_p^{d+\frac{1}{2}}+2 \lambda ^{d+\frac{7}{2}} \lambda_p^{d+\frac{1}{2}}-2 \lambda ^{\frac{d+7}{2}} \lambda_p^{d+\frac{1}{2}}+2 \lambda ^{\frac{d}{2}+2} \lambda_p^{d+\frac{3}{2}}-\lambda ^{d+\frac{1}{2}} \lambda_p^{d+\frac{3}{2}}-2 \lambda ^{d+1} \lambda_p^{d+\frac{3}{2}}-2 \lambda ^{d+\frac{3}{2}} \lambda_p^{d+\frac{3}{2}}\nonumber\\
 &+2 \lambda ^{\frac{d+3}{2}} \lambda_p^{d+\frac{3}{2}}+2 \lambda ^{\frac{d+5}{2}} \lambda_p^{d+\frac{3}{2}}
 \Bigg)\Bigg/ \left(\left(\sqrt{\lambda }+1\right)^3 \sqrt{\lambda_p} \left(\lambda ^{3/2}-\lambda_p\right)\right)\Bigg].\label{eq8}
 \end{align}

Substituting Eqs.~\eqref{eq1},~\eqref{eq2},~\eqref{eq3},~\eqref{eq4},~\eqref{eq5},~\eqref{eq6},~\eqref{eq7}, and ~\eqref{eq8} in Eq.~\eqref{sigmafinal1} we obtain,
 \begin{align}
 \sigma^{(1)}(z)&=\frac{a^2\lambda^{-2d}}{2}\left[2d-1+
 \frac{2\sqrt{\lambda}}{\lambda-1}
 +\frac{\sqrt{\lambda}+\lambda+\lambda^{3/2}}{(\sqrt{\lambda}-1)(\sqrt{\lambda}+1)^3}+\frac{\lambda^2}{a^2(\sqrt{\lambda}-1)(\sqrt{\lambda}+1)^3}-2\sum\limits_{p=2}^d\frac{a_p(\lambda^2-\lambda_p)}{a(1-\frac{\lambda_p}{\lambda})(\sqrt{\lambda}+1)(\lambda^{3/2}-\lambda_p)}\right.\nonumber\\
& -\sum\limits_{p,q=2}^d \frac{a_pa_q(\sqrt{\lambda}-1)(\lambda^3-\lambda_p\lambda_q)}{a^2(1-\frac{\lambda_p\lambda_q}{\lambda^2})(\sqrt{\lambda}+1)(\lambda^{3/2}-\lambda_p)(\lambda^{3/2}-\lambda_q)}
-2\sum\limits_{p=2}^d\frac{a_p(2\lambda^2+2\lambda^{5/2}+\lambda^3-\lambda_p(1+2\sqrt{\lambda}+2\lambda))}{a(1-\frac{\lambda_p}{\lambda})(\sqrt{\lambda}+1)^3(\lambda^{3/2}-\lambda_p)}\nonumber\\
&-\sum\limits_{p,q=2}^d \frac{a_pa_q(\lambda^3(\lambda^{3/2}+\lambda+\sqrt{\lambda}-1)-\lambda^2(\lambda_p+\lambda_q)(1+\sqrt{\lambda})+\lambda_p\lambda_q(1+\sqrt{\lambda}+\lambda-\lambda^{3/2}))        )}{a^2(1-\frac{\lambda_p\lambda_q}{\lambda^2})(\sqrt{\lambda}+1)^3(\lambda^{3/2}-\lambda_p)(\lambda^{3/2}-\lambda_q)}\nonumber\\
&-\frac{2\lambda^{(1-3d)/2}(a(1-3d)\lambda^d-\frac{\lambda}{\lambda-1})}{a^2(\sqrt{\lambda}+1)}-\frac{2\lambda^{(1-d)/2}(1+3d(\sqrt{\lambda}-1))}{a(\lambda-1)}
 -\frac{2\lambda^{-3(d-1)/2}}{a^2(\sqrt{\lambda}+1)^2(\lambda^{3/2}-1)}\nonumber\\
 &-2\sum\limits_{p=2}^d\frac{a_p\lambda^{-3(d-1)/2}\lambda_p^d}{a^2(\lambda-\lambda_p)(\sqrt{\lambda}+1)}\left(1 
 -\frac{(\sqrt{\lambda}-1)\lambda_p}{\lambda^{3/2}-\lambda_p}\right)-\frac{2\lambda^{(1-2d)/2}(1+\sqrt{\lambda}+\lambda)(a(1-2d)\lambda^d-\frac{\lambda}{\lambda-1})}{a^2(\sqrt{\lambda}+1)^3}\nonumber\\
 &-\frac{2\lambda^{(3-2d)/2}}{a^2(\sqrt{\lambda}-1)(\sqrt{\lambda}+1)^4}-\frac{2\sqrt{\lambda}(1+\sqrt{\lambda}+\lambda)(1+2d(\sqrt{\lambda}-1) ) }{a(\sqrt{\lambda}+1)^3(\sqrt{\lambda}-1)}\nonumber\\
 &-2\sum\limits_{p=2}^d\frac{a_p\lambda^{(3-2d)/2}\lambda_p^d(1+\sqrt{\lambda}+\lambda)}{a^2(\lambda-\lambda_p)(\sqrt{\lambda}+1)^3}\left(1 
 -\frac{(\sqrt{\lambda}-1)\lambda_p}{\lambda^{3/2}-\lambda_p}\right) 
 -\frac{2 \lambda ^{2-\frac{3 d}{2}}}{a^2 \left(\sqrt{\lambda }+1\right)^3 \left(\lambda ^2-1\right)}+\frac{2 \left(\lambda +\sqrt{\lambda }+1\right) \lambda ^{1-\frac{d}{2}}}{a \left(\sqrt{\lambda }-1\right) \left(\sqrt{\lambda }+1\right)^4}\nonumber\\
& -\frac{2 \left(-2 \lambda ^{2-\frac{d}{2}}-2 \lambda ^{3-\frac{d}{2}}-\lambda ^{\frac{3}{2}-\frac{d}{2}}-3 \lambda ^{\frac{5}{2}-\frac{d}{2}}-\lambda ^{\frac{7}{2}-\frac{d}{2}}+5 \lambda ^{3/2}+3 \lambda ^{5/2}+\lambda ^3+5 \lambda ^2+3 \lambda +\sqrt{\lambda }\right)}{a \left(\sqrt{\lambda }+1\right)^4 \left(\lambda ^{3/2}-1\right)}\nonumber\\
 & +\frac{2 \lambda ^{1-\frac{3 d}{2}} \left(2 \lambda ^{\frac{d}{2}+1}+3 \lambda ^{\frac{d}{2}+2}+\lambda ^{\frac{d}{2}+3}+\lambda ^{\frac{d+1}{2}}+3 \lambda ^{\frac{d+3}{2}}+2 \lambda ^{\frac{d+5}{2}}-\lambda ^{3/2}-3 \lambda ^{5/2}-\lambda ^{7/2}-2 \lambda ^3-2 \lambda ^2\right)}{a^2 \left(\sqrt{\lambda }+1\right)^4 \left(\lambda ^{3/2}+\lambda ^{5/2}-\lambda -1\right)}\nonumber\\
 &-2\sum\limits_{p=2}^d\frac{a_p\lambda^2}{a^2(\lambda^2-\lambda_p)}\Bigg[\lambda^{-1-2d}\Bigg(
 \lambda ^3 \lambda_p(\sqrt{\lambda } \sqrt{\lambda_p})^{2 d-1}-\lambda\lambda_p^{3/2} \left(\sqrt{\lambda } \sqrt{\lambda_p}\right)^{2 d} -2 \lambda ^{\frac{d}{2}+3} \lambda_p^{d+\frac{1}{2}}-2 \lambda ^{\frac{d}{2}+4} \lambda_p^{d+\frac{1}{2}}-\lambda^{d+2} \lambda_p^{d+\frac{1}{2}}\nonumber\\
 &+2 \lambda^{d+\frac{5}{2}} \lambda_p^{d+\frac{1}{2}}+2 \lambda^{d+3} \lambda_p^{d+\frac{1}{2}}+2 \lambda ^{d+\frac{7}{2}} \lambda_p^{d+\frac{1}{2}}-2 \lambda ^{\frac{d+7}{2}} \lambda_p^{d+\frac{1}{2}}+2 \lambda ^{\frac{d}{2}+2} \lambda_p^{d+\frac{3}{2}}-\lambda ^{d+\frac{1}{2}} \lambda_p^{d+\frac{3}{2}}-2 \lambda ^{d+1} \lambda_p^{d+\frac{3}{2}}-2 \lambda ^{d+\frac{3}{2}} \lambda_p^{d+\frac{3}{2}}\nonumber\\
 &+2 \lambda ^{\frac{d+3}{2}} \lambda_p^{d+\frac{3}{2}}+2 \lambda ^{\frac{d+5}{2}} \lambda_p^{d+\frac{3}{2}}
 \Bigg)\Bigg/ \left(\left(\sqrt{\lambda }+1\right)^3 \sqrt{\lambda_p} \left(\lambda ^{3/2}-\lambda_p\right)\right)\Bigg] +2\sum\limits_{p=2}^d\frac{a_p\lambda^2}{a^2(\lambda^2-\lambda_p)}\frac{\lambda^{-3d/2}(\lambda^{\frac{3}{2}}-1)\lambda_p^{d+1}}{(\sqrt{\lambda}+1)^3(\lambda^{\frac{3}{2}}-\lambda_p)}
 \Bigg].\label{sigmafinal2}
 \end{align}

\begin{figure}
\includegraphics[width=0.7\columnwidth]{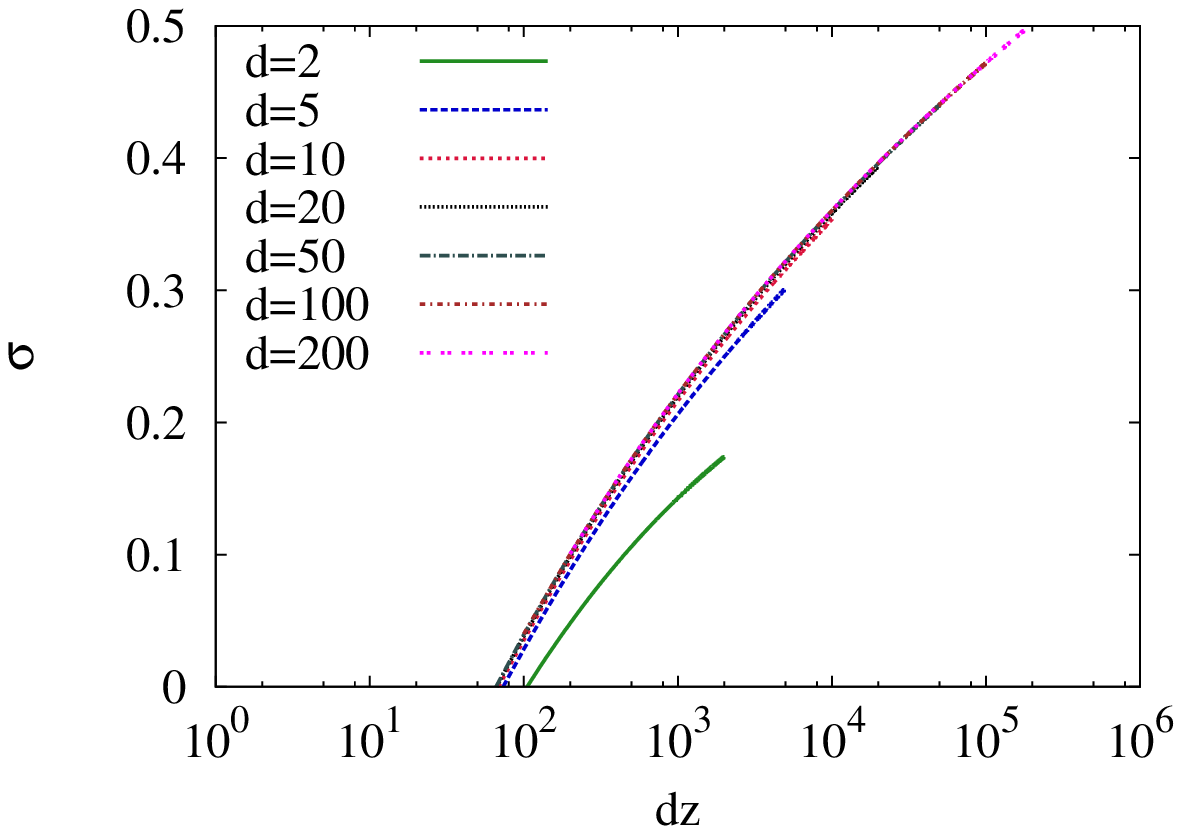}
\caption{Variation of interfacial tension $\sigma(z)$ truncated at order $z_D$ with $dz$} 
\label{fig:sigma}
\end{figure}

The variation interfacial tension, $\sigma(z)=\sigma^{(0)}(z)+z_D\sigma^{(1)}(z)$,  with $z$
is shown in Fig.~\ref{fig:sigma}. The function decreases monotonically with decreasing $z$ till it 
becomes zero at $z_c$. Near $z_c$, it varies as $|z-z_c|$.

\section{\label{D}Calculation of $\rho(d)$ with one defect}

The bulk density may be calculated from $ \Omega_{ee}$ (Eq.~\eqref{omegaee4}),
\be
\Omega_{ee}=[\omega_0(N)]^{M/2}\left(1+z_D\frac{M}{2} \frac{a^2}{\lambda^{2d}}[(N-d+1)+2f(0)]+O(z_D^2)\right).
\ee
At thermodynamic limit, free energy per site is thus given by,
\be
f=-\frac{1}{2}\left[\ln \lambda+z_Da^2\lambda^{-2d}\right]+O(z_D^2).
\ee
Density of rectangles, $\rho(d)$ is defined as
\be
\rho(d)=-dz\frac{d}{dz}f,
\ee
which may be simplified to
\be
\rho(d)=1-\frac{a}{\lambda}+z_Da^2\lambda^{-1-2d}\left[2a(d(a+1)-1)-\lambda(d(2a+1)-2) \right]+O(z_D^2).
\ee

\section{\label{E}Calculation of contribution from overhang}

A configuration without any overhangs correspond to a PDSAW with only left, right and down steps allowed. 
We denote weight for vertical downward step as $\mathbb{D}$, upward step as $\mathbb{U}$, left step as $\mathbb{L}$ and right step as $\mathbb{R}$.  The steps will be denoted by $D,U,L$ and $R$ respectively, where
\begin{align}
 \mathbb{D}=az\lambda^{-d},\label{dwt}\\
 \mathbb{R}=\mathbb{L}=\lambda^{-1/2}.\label{rwt}
\end{align}

A PDSAW consists of an  arbitrary concatenation of substrings  $D$,  $DR$, $DR^2$,$\ldots$,  $DL$, $DL^2$,$\ldots$ etc. Hence 
the PDSAW has a formal generating function
\begin{equation}\label{genfun1}
 \wp=\frac{1}{1-D-DR^*-DL^*},
\end{equation}
where 
\begin{equation}
 R^*=R+R^2+R^3+\ldots=\frac{R}{1-R}.
\end{equation}
and a similar equation for $L^*$. Correspondingly, the equation for the evaluation of this 
generating function is 
\be
W = \frac{1}{1 - \mathbb{D} - \mathbb{DR}^* - \mathbb{DL}^*}.
\ee

The criticality condition when the generating function diverges, then becomes
\begin{equation}\label{critic}
 \mathbb{D}(1+\mathbb{R}^*+\mathbb{L}^*)=1.
\end{equation}
which is the condition of interfacial tension $\sigma^{(0)}(z)$ becoming zero. Using Eq. (10) of main text, this may be simplified to give 
\begin{equation}
 a ( 1 + 1/\sqrt{\lambda})^2 =1.
\end{equation}

Now consider interface configurations with overhangs of height 1. 
The simplest sequence of steps corresponding to an overhang is $D$$R^a$$U$$R^b$, 
with $a,b \geq d$. Let us define
\be
W_R = \sum_{a,b = d}^{\infty}  R^a U R^b D
\ee
and arbitrary concatenation of such overhangs at same height is given by
\be
W_R^* = W_R + W_R^2 + W_R^3 +... = W_R/[ 1 - W_R]
\ee
We define $W_L$, and $W_L^*$ similarly, with $R$ replaced by $L$, corresponding to overhangs to the left of the interface. Given an overhang configuration with a given values of $a$ and $b$, it is easy to see that corresponding sum over allowed rectangle configurations to the left and right of interface gives a relative weight $ z ^2 \omega(a-d) \omega(b-d) \lambda^{-3(a+b)/2}$.  Then summing over $a$, and $b$, we get 
\begin{equation}
 \mathbb{W}_R=\sum\limits_{a=d}^{\infty}\sum\limits_{b=d}^{\infty}z^2\lambda^{-\frac{3}{2}(a+b)}\omega_0(a-d)\omega_0(b-d),
 \end{equation}
which, using Eq. (12) of main text  can be simplified to
\begin{align}
 \mathbb{W}_R&=[z\lambda^{-\frac{3d}{2}}]^2\left[\sum\limits_{n=0}^{\infty}\omega_0(n) \lambda^{-\frac{3n}{2}}
 \right]^2,\\
 &=\left[\frac{z\lambda^{-\frac{3d}{2}}}{1-\lambda^{-3/2}-(1-\lambda^{-1})\lambda^{-d/2}} \right]^2.
\end{align}

A general configuration with overhangs of height at most $1$ is an arbitrary concatenation of subsequences $D, DR^*, DL^*, X_L$ and $X_R$, where  
\begin{eqnarray}
X_R = D\left(\left(\sum_{n \geq 0}   L^n D R^n \right)+ R^* D\right) W_R^* ( 1 + R^*),\\
X_L = D\left(\left(\sum_{n \geq 0}   R^n D L^n \right)+ L^* D\right) W_L^* ( 1 + L^*). \\ 
\end{eqnarray}
Thus the criterion for divergence of the generating function Eq.~\eqref{critic} gets modified to
\begin{equation}
 \mathbb{D}\left(1+2\mathbb{R}^*+2\left(\frac{1}{1-\mathbb{LR}}+\mathbb{R}^*\right)\frac{\mathbb{DW}_R(1+\mathbb{R}^*)}{(1- \mathbb{W}_R)}\right)=1,
\end{equation}
which can be simplified to
\begin{equation}
 \frac{\lambda-1}{d(\lambda-1)+1}\left(\frac{\sqrt{\lambda}+1}{\sqrt{\lambda}-1}+\frac{2\sqrt{\lambda}(1+\sqrt{\lambda}+\lambda)\mathbb{W}_R}{(d(\lambda-1)+1)(\sqrt{\lambda}-1)(1- \mathbb{W}_R)}\right)=1.
\end{equation}